\documentstyle[epsf, rotate]{aipproc}

\begin{document}
\title{Leptonic Jet Models of Blazars:\\
Broadband Spectra and Spectral Variability}

\author{Markus B\"ottcher$^*$}
\address{$^*$Space Physics and Astronomy Department \\
Rice University, MS 108 \\
6100 S. Main Street \\
Houston, TX 77005 - 1892 \\
USA}

\maketitle

\begin{abstract}
The current status of leptonic jet models for
gamma-ray blazars is reviewed. Differences 
between the quasar and BL-Lac subclasses of 
blazars may be understood in terms of the
dominance of different radiation mechanisms 
in the gamma-ray regime. Spectral variability
patterns of different blazar subclasses
appear to be significantly different and
require different intrinsic mechanisms
causing gamma-ray flares. As examples,
recent results of long-term multiwavelength 
monitoring of PKS 0528+134 and Mrk 501 
are interpreted in the framework of leptonic 
jet models. A simple quasi-analytic toy model
for broadband spectral variability of blazars
is presented.
\end{abstract}

\section*{Introduction}

Recent high-energy detections and simultaneous broadband 
observations of blazars, determining their spectra and spectral 
variability, are posing strong constraints on currently popular 
jet models of blazars. 66 blazars have been detected by EGRET at
energies above 100~MeV \cite{hartman99a}, the two nearby
high-frequency peaked BL~Lac objects (HBLs) Mrk~421 and 
Mrk~501 are now multiply confirmed sources of multi-GeV 
-- TeV radiation \cite{punch92,petry96,quinn96,brad97}, 
and the TeV detections of PKS~2155-314 \cite{chadwick99} 
and 1ES~2344+514 \cite{catanese98} are awaiting confirmation. 
Most EGRET-detected blazars exhibit rapid variability
\cite{mukherjee97}, in some cases on intraday and even 
sub-hour (e. g., \cite{gaidos96}) timescales, where generally 
the most rapid variations are observed at the highest photon 
frequencies.

The broadband spectra of blazars consist of at least two 
clearly distinct spectral components. The first one extends
in the case of flat-spectrum radio quasars (FSRQs) from radio 
to optical/UV frequencies, in the case of HBLs up to soft and
even hard X-rays, and is consistent with non-thermal synchrotron
radiation from ultrarelativistic electrons. The second spectral
component emerges at $\gamma$-ray energies and peaks at several
MeV -- a few GeV in most quasars, while in the case of some 
HBLs the peak of this component appears to be located at TeV 
energies. 

The bolometric luminosity of EGRET-detected quasars and some
low-frequency peaked BL~Lac objects (LBLs) during flares is
dominated by the $\gamma$-ray emission. If this emission were 
isotropic, it would correspond to enormous luminosities (up to 
$\sim 10^{49}$~erg~s$^{-1}$) which, in combination with the
short observed timescales (implying a small size of the 
emission region) would lead to a strong modification of the
emissivity spectra by $\gamma\gamma$ absorption, in contradiction 
to the observed smooth power-laws at EGRET energies. This has 
motivated the concept of relativistic beaming of radiation 
emitted by ultrarelativistic particles moving at relativistic 
bulk speed along a jet (for a review of these arguments, see 
\cite{schlickeiser96}). While it is generally accepted that 
blazar emission originates in relativistic jets, the radiation 
mechanisms responsible for the observed $\gamma$ radiation are 
still under debate. It is not clear yet whether in these jets 
protons are the primarily accelerated particles, which then 
produce the $\gamma$ radiation via photo-pair and photo-pion 
production, followed by $\pi^0$ decay and synchrotron emission 
by secondary particles (e. g., \cite{mannheim93}), or electrons 
(and positrons) are accelerated directly and produce $\gamma$-rays 
in Compton scattering interactions with the various target photon 
fields in the jet \cite{mg85,dermer92,sikora94,boettcher97}. 

In this review, I will describe the current status of blazar
models based on leptons (electrons and/or pairs; in the following, 
the term ``electrons'' refers to both electrons and positrons) as 
the primary constituents of the jet which are responsible for the 
$\gamma$-ray emission. Hadronic jet models are discussed in a 
separate paper by J. Rachen \cite{rachen99}. In Section 2, I 
will give a description of the model and discuss the different 
$\gamma$-ray production mechanisms and their relevance for 
different blazar classes. In Section 3, I will review recent 
progress in understanding intrinsic differences between different 
blazar classes and present state-of-the-art model calculations, 
using a leptonic jet model, to undermine the general theoretical 
concept. In Section 4, I will discuss broadband spectral variability
of individual blazars and their interpretation in the framework
of leptonic jet models. A simple quasi-analytical toy model for
blazar broadband spectral variability will be presented in
section 5.

\section*{Model description and radiation mechanisms}

\begin{figure}
\epsfysize=6cm
\hskip 3cm \epsffile[0 0 400 500]{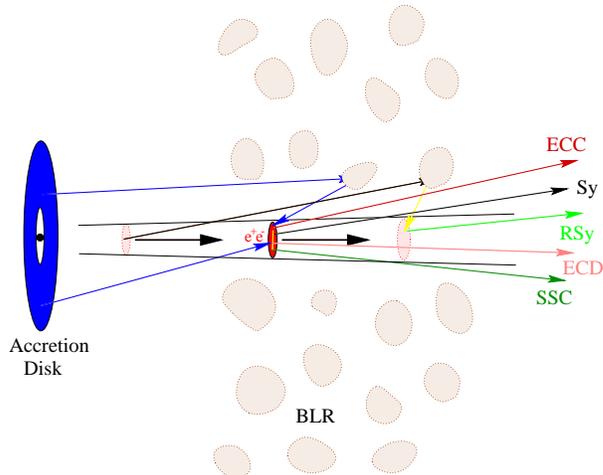}
\bigskip
\caption[]{Illustration of the model geometry and the relevant 
$\gamma$ radiation mechanisms for leptonic jet models.}
\label{sketch}
\end{figure}

The basic geometry of leptonic blazar jet models is illustrated
in Fig. \ref{sketch}. At the center of the AGN, an accretion
disk around a supermassive, probably rotating, black hole is
powering a relativistic jet. Along this pre-existing jet structure,
occasionally blobs of ultrarelativistic electrons are ejected at 
relativistic bulk velocity. 

The electrons are emitting synchrotron radiation, which will 
be observable at IR -- UV or even X-ray frequencies, and hard 
X-rays and $\gamma$-rays via Compton scattering processes. Possible 
target photon fields for Compton scattering are the synchrotron photons
produced within the jet (the SSC process, \cite{mg85,maraschi92,bloom96}),
the UV -- soft X-ray emission from the disk --- either entering the
jet directly (the ECD [External Comptonization of Direct disk radiation]
process; \cite{dermer92,ds93}) or after reprocessing at the broad line
regions or other circumnuclear material (the ECC [External Comptonization
of radiation from Clouds] process; \cite{sikora94,bl95,dss97}), or jet
synchrotron radiation reflected at the broad line regions (the RSy
[Reflected Synchrotron] mechanism; \cite{gm96,bednarek98,bd98}).

The relative importance of these components may be estimated by
comparing the energy densities of the respective target photon
fields. Denoting by $u'_B$ the co-moving energy density of the 
magnetic field, the energy density of the synchrotron radiation
field, governing the luminosity of the SSC component, may be 
estimated by $u'_{sy} \approx u'_B \, \tau_T \, \gamma_e^2$,
where $\tau_T = n'_{e, B} \, R'_B \, \sigma_T$ is the Thomson depth 
of the relativistic plasma blob and $\gamma_e$ is the average
Lorentz factor of electrons in the blob. The SSC spectrum exhibits
a broad hump without strong spectral break, peaking around 
$\langle\epsilon\rangle_{SSC} \approx (B' / B_{cr}) \, D \,
\gamma_e^4 \approx \langle\epsilon\rangle_{sy} \, \gamma_e^2$, 
where $B'$ is the co-moving magnetic field, $B_{cr} = 4.414 
\cdot 10^{13}$~G, and $D = \left( \Gamma \, [1 - \beta_{\Gamma}
\cos\theta_{obs}] \right)^{-1}$ is the Doppler factor associated
with the bulk motion of the blob. Throughout this paper, all 
photon energies are described by the dimensionless quantity 
$\epsilon = h \nu / (m_e c^2)$. 

If the blob is sufficiently far from the central engine of the AGN
so that the accretion disk can be approximated as a point source of
photons, its photon energy density (in the co-moving frame) is 
$u'_D \approx L_D / (4 \pi \, z^2 \, c \, \Gamma^2)$, where $L_D$
is the accretion disk luminosity, and $z$ is the height of the blob
above the accretion disk. The ECD spectrum can exhibit a strong
spectral break, depending on the existence of a low-energy cutoff
in the electron distribution function, and peaks at $\langle\epsilon
\rangle_{ECD} \approx \langle\epsilon\rangle_D \, (D / \Gamma) \,
\gamma_e^2$, where $\langle\epsilon\rangle_D$ is the average photon
energy of the accretion disk radiation (typically of order $10^{-5}$
for Shakura-Sunyaev type accretion disks \cite{shakura73} around black
holes of $\sim 10^8$ -- $10^{10} \, M_{\odot}$).

Part of the accretion disk and the synchrotron radiation will be
reprocessed by circumnuclear material in the broad line region
and can re-enter the jet. Since this reprocessed radiation is 
nearly isotropic in the rest-frame of the AGN, it will be strongly
blue-shifted into the rest-frame of the relativistically moving
plasma blob. Thus, assuming that a fraction $a_{BLR}$ of the radiation
is rescattered into the jet trajectory, we find for the energy density 
of rescattered accretion disk photons: $u'_{ECC} \approx L_D \, a_{BLR}
\, \Gamma^2 / (4 \pi \, \langle r \rangle_{BLR}^2 \, c)$, where
$\langle r \rangle_{BLR}$ is the average distance of the BLR material
from the central black hole. The ECC photon spectrum peaks around
$\langle\epsilon\rangle_{ECC} \approx \langle\epsilon\rangle_D \,
D \, \Gamma \, \gamma_e^2 \approx \langle\epsilon\rangle_{ECD} \, 
\Gamma^2$. 

For the synchrotron mirror mechanism, additional constraints due to
light travel time effects need to be taken into account in order to
estimate the reflected synchrotron photon energy density (for a
detailed discussion see \cite{bd98}), which is well approximated
by $u'_{RSy} \approx u'_{sy} \, 4 \,\Gamma^3 \, a_{BLR} \, (R'_B 
/ \Delta r_{BLR}) \, ( 1 - 2 \, \Gamma \, R'_B / z)$, where $\Delta
r_{BLR}$ is a measure of the geometrical thickness of the broad line 
region. Similar to the SSC spectrum, the RSy spectrum does not show
a strong spectral break. It peaks around $\langle\epsilon\rangle_{RSy}
\approx (B' / B_{cr}) \, D \, \Gamma^2 \, \gamma_e^4 \approx \langle
\epsilon \rangle_{SSC} \, \Gamma^2$. 

\begin{figure}
\epsfysize=5cm
\hskip 3cm \rotate[r]{\epsffile[100 0 550 400]{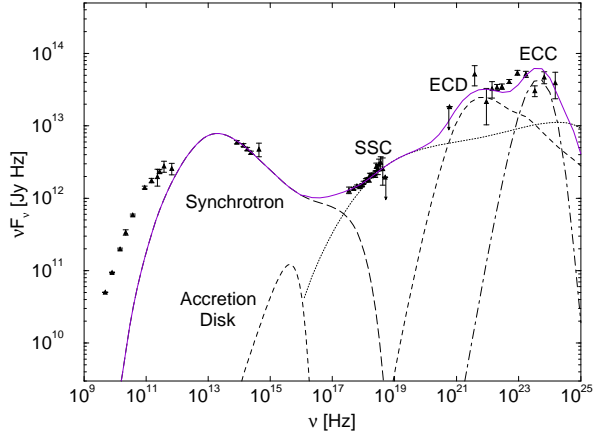}}
\bigskip
\caption[]{Fit to the simultaneous broadband spectrum of the FSRQ
3C279 during its very bright $\gamma$-ray flare in September 1991.
See \cite{hartman99b} for model parameters.}
\label{3c279_fit}
\end{figure}

\section*{Trends between different blazar classes}

There appears to be a more or less continuous sequence in the broadband
spectral properties of blazars, ranging from FSRQs over LBLs to HBLs, which 
was first presented in a systematic way in \cite{fossati97}. While in
FSRQs the synchrotron and $\gamma$-ray peaks are typically located at 
infrared and MeV -- GeV energies, respectively, they are shifted towards
higher frequencies in BL~Lacs, occurring at medium to even hard X-rays
and at multi-GeV -- TeV energies in some HBLs. The bolometric luminosity
of FSRQs is --- at least during $\gamma$-ray high states --- strongly 
dominated by the $\gamma$-ray emission, while in HBLs the relative
power outputs in synchrotron and $\gamma$-ray emission are comparable.

\begin{figure}
\epsfysize=5cm
\hskip 3cm \rotate[r]{\epsffile[100 0 550 400]{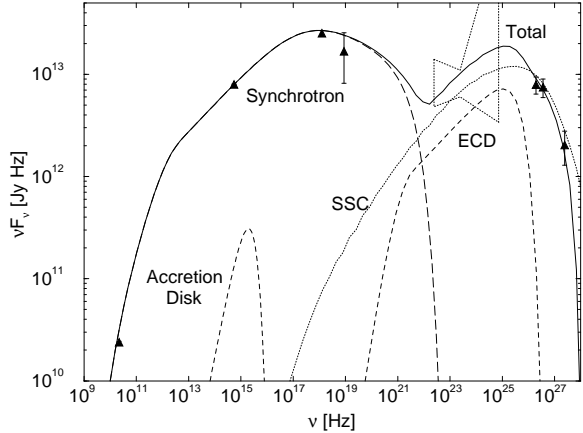}}
\bigskip
\caption[]{Fit to a weekly-averaged broadband spectrum of the
extreme HBL Mrk~501 during a high $\gamma$-ray state centered
on MJD~50564 in 1997. The dotted ``bow-tie'' curve indicates
the highest flux ever measured by EGRET, which provides an
upper limit for the possible contribution of external Comptonization.
See \cite{petry99} for model parameters.}
\label{mrk501_fit}
\end{figure}

Detailed modeling of several blazars has indicated that this
sequence appears to be related to the relative contribution of
the external Comptonization mechanisms ECD and ECC to the $\gamma$-ray
spectrum. While most FSRQs are successfully modelled with external
Comptonization models (e. g., \cite{dss97,sambruna97,muk99,hartman99b}), 
the broadband spectra of HBLs are consistent with pure SSC models (e. g., 
\cite{mk97,pian98,petry99}). BL~Lacertae, a LBL, appears to be 
intermediate between these two extremes, requiring an external 
Comptonization component to explain the EGRET spectrum 
\cite{madejski99,bb99}. Figs. \ref{3c279_fit} and \ref{mrk501_fit} 
illustrate detailed modeling results of two objects located at 
opposite ends of this sequence of blazars, using the jet radiation 
transfer code described in \cite{boettcher97,bb99}.

A physical interpretation of this sequence in the framework of a
unified jet model for blazars was given in \cite{ghisellini98}.
Assume that the average energy of electrons, $\gamma_e$, is
determined by the balance of an energy-independent acceleration
rate $\dot\gamma_{acc}$ and radiative losses, $\dot\gamma_{rad}
\approx - (4/3) \, c \, \sigma_T \, (u' / m_e c^2) \, \gamma^2$,
where the target photon density $u'$ is the sum of the sources
intrinsic to the jet, $u'_B + u'_{sy}$ plus external photon
sources, $u'_{ECD} + u'_{ECC} + u'_{RSy}$. The average electron
energy will then be $\gamma_e \propto (\dot\gamma_{acc} / u')^{1/2}$.
If one assumes that the properties determining the acceleration
rate of relativistic electrons do not vary significantly between 
different blazar subclasses, then an increasing energy density
of the external radiation field will obviously lead to a
stronger radiation component due to external Comptonization,
but also to a decreasing average electron energy $\gamma_e$,
implying that the peak frequencies of both spectral components
are displaced towards lower frequencies.

\section*{Spectral variability of blazars}

Between flaring and non-flaring states, blazars show very distinct
spectral variability. Not only does the emission at the highest
frequencies generally vary on the shortest time scales, but also
the flaring amplitudes are significantly different among different
wavelength bands. FSRQs often show spectral hardening of their
$\gamma$-ray spectra during $\gamma$-ray flares (e. g., 
\cite{collmar97,hartman96,wehrle98}), and the flaring amplitude
in $\gamma$-rays is generally larger than in all other wavelength
bands. The concept of multi-component $\gamma$-ray spectra of 
quasars, as first suggested for PKS~0528+134 in \cite{collmar97}, 
offers a plausible explanation for this spectral variability due 
to the different beaming patterns of different radiation mechanisms,
as pointed out in \cite{dermer95}. This has been applied to 
PKS~0528+134 in \cite{bc98} and \cite{muk99}. 

\begin{figure}
\epsfysize=5cm
\hskip 3cm \rotate[r]{\epsffile[100 0 550 400]{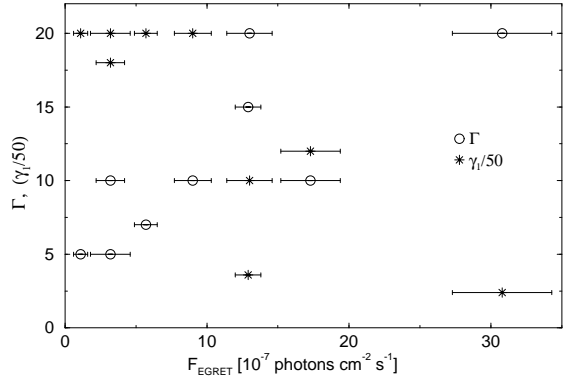}}
\bigskip
\caption[]{The dependence of the fit parameters $\Gamma$ and
$\gamma_1$ (low-energy cut-off of the electron distribution)
on the EGRET flux for fits to simultaneous broadband spectra
of PKS~0528+134 (see \cite{muk99}).}
\label{0528_parameters}
\end{figure}

The results of \cite{muk99} indicate that $\gamma$-ray flaring
states of PKS~0528+134 are consistent with an increasing bulk
Lorentz factor $\Gamma$ of ejected jet material, while at the same 
time the low-energy cutoff $\gamma_1$ of the electron distribution 
injected into the jet is lowered. This is in agreement with the
physical picture that due to an increasing $\Gamma$, the quasi-isotropic
external photon field is more strongly Lorentz boosted into the blob
rest frame, leading to stronger external Compton losses, implying a
lower value of $\gamma_1$. The external Compton $\gamma$-ray components
depend much more strongly on the bulk Lorentz factor than the synchrotron
and SSC components do. This leads naturally to a hardening of the
$\gamma$-ray spectrum, if the SSC mechanisms plays an important or
even dominant role in the X-ray --- soft $\gamma$-ray regime, while
external Comptonization is the dominant radiation mechanism at higher
$\gamma$-ray energies. The results of this study on PKS~0528+134
are discussed in more detail in \cite{muk99p}.

While this flaring mechanism is plausible for FSRQs, short-timescale, 
correlated X-ray and $\gamma$-ray flares of the HBLs Mrk~421 and 
Mrk~501 \cite{mk97,pian98} and synchrotron flares of other HBLs (e. g., 
PKS~2155-304, \cite{georg98,kataoka99}) have been explained successfully 
in the context of SSC models where flares are related to an increase of 
the maximum electron energy, $\gamma_2$, and a hardening of the 
electron spectrum. How the spectral evolution in synchrotron flares
can be used to constrain the magnetic field and the physics of
injection and acceleration of relativistic pairs, has been described
in detail in the previous talk by R. Sambruna \cite{sambruna99}.

\begin{figure}
\epsfysize=5cm
\hskip 2.5cm \rotate[r]{\epsffile[100 0 550 400]{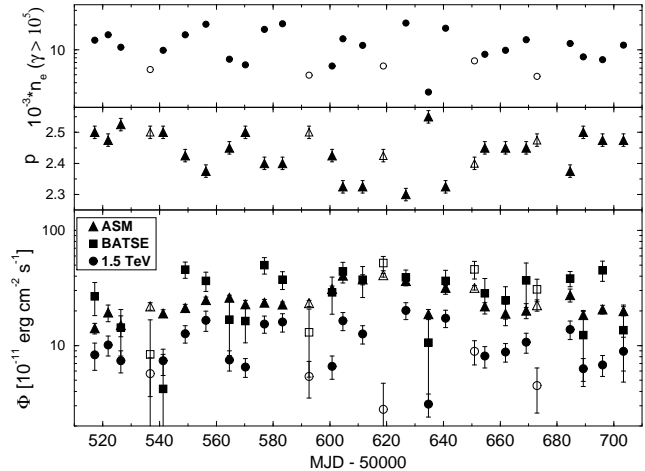}}
\bigskip
\caption[]{Temporal variation of the fit parameters $n_e (\gamma
> 10^5)$ (density of high-energy electrons) and $p$ (spectral
index of injected electron distribution) compared to the weekly
averaged light curves from RXTE ASM, BATSE and HEGRA (see 
\cite{petry99}).}
\label{mrk501_parameters}
\end{figure}

Comparing detailed spectral fits to weekly averaged broadband spectra
of Mrk~501 \cite{petry99} over a period of 6 months, we have found 
that TeV and hard X-ray high states on intermediate timescales are 
consistent with a hardening of the electron spectrum (decreasing 
spectral index) and an increasing number density of high-energy 
electrons, while the value of $\gamma_2$ has only minor influence 
on the weekly averaged spectra. Fig. \ref{mrk501_parameters} shows
how the spectral index of the injected electron distribution and 
the density of high-energy electrons resulting from our fits are
varying in comparison to the RXTE ASM, BATSE, and HEGRA 1.5~TeV
light curves. For a more detailed discussion of this analysis
see \cite{boettcher99}.

These variability studies seem to indicate that due to the different
dominant $\gamma$ radiation mechanisms in quasars and HBLs also the
physics of $\gamma$-ray flares and extended high states is considerably 
different. While in FSRQs the $\gamma$-ray emission and its flaring
behavior appears to be dominated by conditions of the external radiation
field, this influence is unimportant in the case of HBLs where emission
lines are very weak or absent, implying that the BLR might be very
dilute, leading to a very weak external radiation field, which becomes
negligible compared to the synchrotron radiation field intrinsic to
the jet. 

\section*{A toy model for spectral variability}

On the basis of the estimates of the photon energy densities of the
various radiation fields and the peak energies of the diverse radiation
components given in Section 2, one can develop a very simple, quasi-analytic 
toy model which allows us to study the influence of parameter variations 
on the predicted broadband spectrum of a blazar. For construction of this
toy model, I assume that the magnetic field within the jet is in equipartition
with the ultrarelativistic electrons, and that the light-travel time effects
affecting the efficiency of the synchrotron mirror mechanism can be 
parametrized by a correction factor $f_{ltt} \lesssim 0.1$ so that
$u'_{RSy} \approx u'_{sy} \, a_{BLR} \, \Gamma^3 \, f_{ltt}$. Then,
the entire broadband spectrum, accounting for all synchrotron and
inverse-Compton components, is determined by 9 parameters: $\Gamma$,
$\theta_{obs}$, $\dot\gamma_{acc}$, $L_D$, $\langle\epsilon\rangle_D$,
$a_{BLR}$, the scale height $z$ of energy dissipation in the jet,
$R'_B$, and $\tau_T$. In most cases, several of these parameters can
be constrained by independent observations. The radiation spectra
of each individual component are approximated by double power-laws
with a smooth transition.

\begin{figure}
\epsfysize=5cm
\hskip 3cm \rotate[r]{\epsffile[100 0 550 400]{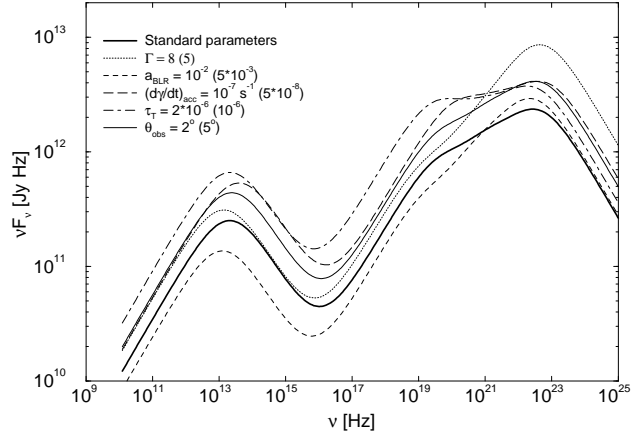}}
\bigskip
\caption[]{Thick solid curve: Toy model calculation representative 
of the broadband spectrum of PKS~0528+134. Parameters: $\Gamma = 5$, 
$\theta_{obs} = 5^o$, $\dot\gamma_{acc} = 5 \cdot 10^{-8}$~s$^{-1}$, 
$L_D = 6 \cdot 10^{46}$~erg~s$^{-1}$, $\langle\epsilon\rangle_D = 
10^{-5}$, $a_{BLR} = 5 \cdot 10^{-3}$, $z = 2 \cdot 10^{17}$~cm, 
$R'_B = 3 \cdot 10^{16}$~cm, $\tau_T = 10^{-6}$. For the other curves,
a single parameter, as indicated by the label, has been changed
(value of the original calculation in parantheses).}
\label{toy_0528}
\end{figure}

The thick solid curve in Fig. \ref{toy_0528} shows a toy model 
calculation with the location of peak frequencies and the relative 
contributions of the $\gamma$ radiation components as found in 
our fits to PKS~0528+134 based on detailed simulations 
\cite{muk99}. The hard X-ray to soft $\gamma$-ray spectrum below
$\sim 1$~MeV is dominated by the SSC mechanism, while at higher
energies, the ECC mechanism is dominant. The other curves in Fig.
\ref{toy_0528} indicate the effect of single parameter changes
on the broadband spectrum which could be thought of as the cause 
of flares at $\gamma$-ray energies. 

From Fig. \ref{toy_0528}, one can see that an increasing bulk 
Lorentz factor leads to a strong flare at $\gamma$-ray energies, 
while only moderate flaring at infrared and X-ray frequencies 
results (dotted curve). A shift of the synchrotron peak towards 
lower frequencies is predicted \cite{boe99b}. If the BLR albedo 
$a_{BLR}$ increases (short-dashed curve), the result is a slight 
increase in the power output at high-energy $\gamma$-rays, 
while due to enhanced external-Compton cooling the flux in the 
synchrotron and SSC components even decreases. An increased 
acceleration rate $\dot\gamma_{acc}$ (long-dashed curve), leads to 
a strong synchrotron and SSC flare, where the largest variability 
amplitude is expected at MeV energies, while only moderate variability 
at higher frequencies is predicted. If the density of relativistic
electrons in the blob increases during a flare (dot-dashed curve), 
the variability amplitude is again predicted to be largest at MeV 
energies and the peak frequencies of all components are expected 
to be shifted slightly towards lower frequencies. A decreasing
observing angle --- which could be a consequence of a bending
jet --- leads to a shift of the entire broadband spectrum 
towards higher fluxes and slightly higher peak frequencies.

Obviously, definite conclusions can not be drawn from this
simplistic toy-model analysis. However, it may support previous
results that an increasing bulk Lorentz factor is a viable and
plausible explanation for the spectral variability observed in 
PKS~0528+134 and possibly also in other FSRQs. 

\begin{figure}
\epsfysize=4.5cm
\hskip 3cm \rotate[r]{\epsffile[100 0 550 400]{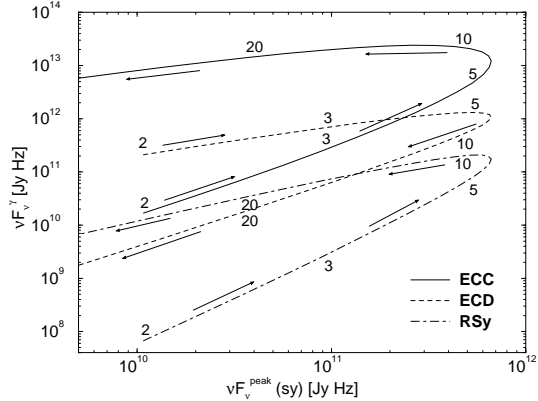}}
\bigskip
\caption[]{Variation of the power output in synchrotron and the external
Comptonization components if the bulk Lorentz factor $\Gamma$ is the
only parameter changing during a flare. The numbers along the curves
are the respective values of $\Gamma$ at that point, and the arrows
indicate the sense of evolution if $\Gamma$ is increasing.}
\label{Gamma_var}
\end{figure}

Fig. \ref{Gamma_var} illustrates the variation of the power output 
in the different external Compton $\gamma$-ray components with
respect to the synchrotron component, if $\Gamma$ is the only parameter
changing during a $\gamma$-ray flare. Most notably, a very steep
relation between $\nu F_{\nu}^{peak} (sy)$ and $\nu F_{\nu}^{peak}
(ECC)$ is predicted for values of the Lorentz factor close to the
critical $\Gamma$, at which the observer is looking at the superluminal
angle. This dependence can be significantly steeper than quadratic,
$\Delta \nu F_{\nu}^{peak} (sy) \propto \left[ \Delta \nu F_{\nu}^{peak} 
(ECC) \right]^{\alpha}$ with $\alpha > 2$, which has recently been 
observed in the prominent 1996 flare of 3C279 \cite{wehrle98}.

\end{document}